\documentclass[12pt,usenames,dvipsnames]{article}

\usepackage{latexsym}
\usepackage{amssymb,amsfonts,amsmath}
\usepackage{graphicx} 
\usepackage{indentfirst}
\usepackage{bbm}
\usepackage{amssymb}
\usepackage{verbatim}
\usepackage{amsmath, amsthm,amssymb}
\usepackage{mathrsfs}
\usepackage{hyperref}
\usepackage{amsfonts}
\usepackage{dsfont}
\usepackage{cite}
\usepackage{xcolor}
\usepackage{enumerate}
\usepackage{cleveref}
\parskip=\medskipamount

\newcommand {\cD}{{\cal D}}
\newcommand {\cE}{{\cal E}}

\newcommand {\cL}{{\cal L}}

\newcommand {\cN}{{\cal N}}

\def\a{\alpha}

\def\d{\delta}

\def\f{\phi}
\def\g{\gamma}

\def\j{\psi}

\def\l{\lambda}
\def\m{\mu}
\def\n{\nu}
\def\o{\omega}

\def\q{\theta}
\def\r{\rho}
\def\s{\sigma}

\def\L{\Lambda}

\def\U{\Upsilon}

\def\rd{{\rm d}}

\newcommand{\ad}{{\dot{\alpha}}}                           %new
\newcommand{\pa}{\partial}                           %new
\newcommand{\hf}{\frac12}
\newcommand{\vf}{\varphi}
\newcommand{\be}{\begin{equation}}
\newcommand{\ee}{\end{equation}}
\newcommand{\bea}{\begin{eqnarray}}
\newcommand{\eea}{\end{eqnarray}}
\newcommand{\non}{\nonumber}
\def\double #1{#1{\hbox{\kern-2pt $#1$}}}

\newif\ifdtup

\newcommand{\bsubeq}{\begin{subequations}}
\newcommand{\esubeq}{\end{subequations}}
\numberwithin{equation}{section}

\newcommand{\sU}{\mathsf{U}}

\let\l\relax
\let\r\relax
\newcommand{\l}{\left}
\newcommand{\r}{\right}
\newcommand{\axd}{\psi^2 -2 \psi  S - P^2}

\newcommand{\wmmbd}{\psi^{2}-2 T \psi  \cosh{\gamma}+T^{2}  }
\newcommand{\er}{\eqref}
\newcommand{\td}{\tilde}

\begin{document}

\begin{titlepage}
\begin{flushright}
November, 2025 \\
\end{flushright}
\vspace{5mm}

\begin{center}
{\Large \bf 
Generalisations of the Russo-Townsend formulation 
}
\end{center}

\begin{center}

{\bf Sergei M. Kuzenko and Jonah Ruhl} \\
\vspace{5mm}

\footnotesize{
{\it Department of Physics M013, The University of Western Australia\\
35 Stirling Highway, Perth W.A. 6009, Australia}}  
~\\
\vspace{2mm}
~\\
Email: \texttt{ 
sergei.kuzenko@uwa.edu.au, 23390035@student.uwa.edu.au}\\
\vspace{2mm}

\end{center}

\begin{abstract}
\baselineskip=14pt
As a generalisation of the recent construction by Russo and Townsend, we propose a new approach to generate $\mathsf{U}(1)$ duality-invariant models for nonlinear electrodynamics. It is based on the use of two building blocks:
(i) a fixed (but otherwise arbitrary) model for self-dual nonlinear electrodynamics with Lagrangian $L(F_{\mu\nu};g)$ depending on a duality-invariant parameter $g$; and (ii) an arbitrary potential $W(\psi)$, with $\psi$ an auxiliary scalar field. It turns out that the model $\mathfrak{L}(F_{\mu\nu};\psi) = L(F_{\mu\nu};\psi) + W(\psi)$ leads to a self-dual theory for nonlinear electrodynamics upon elimination of $\psi$. As an illustration, we work out two examples in which the seed Lagrangian $L(F_{\mu\nu};g)$ corresponds to the Born-Infeld model and two particular potentials $W(\psi)$ are chosen such that integrating out $\psi$  gives: (i)  the ModMaxBorn theory; and (ii) the ModMax theory. We also briefly discuss  supersymmetric generalisations of the proposed formulation.
\end{abstract}
\vspace{10mm}

\begin{flushright}
{\it To the memory of Kelly Stelle}
\end{flushright}

\vfill

\vfill
\end{titlepage}

\newpage
\renewcommand{\thefootnote}{\arabic{footnote}}
\setcounter{footnote}{0}

\tableofcontents{}
\vspace{1cm}
\bigskip\hrule

\allowdisplaybreaks

\section{Introduction}

Recently, Russo and Townsend \cite{Russo:2025fuc} proposed a new formulation for self-dual nonlinear electrodynamics \cite{GZ1, BB, GR1,GR2,GZ2,GZ3}.\footnote{Interesting applications of the Russo-Townsend approach have appeared in \cite{Babaei-Aghbolagh:2025uoz}.}
Their starting point was the ModMax theory proposed by Bandos, Lechner, Sorokin and Townsend
\cite{BLST} and soon re-derived, in a simpler setting, by Kosyakov \cite{Kosyakov}:\footnote{The ModMax theory was also re-derived in \cite{K21} using the Ivanov-Zupnik auxiliary-field formulation \cite{IZ_N3, IZ1, IZ2} for   $\mathsf{U}(1)$ duality-invariant models for nonlinear electrodynamics. }
\begin{align}\label{ModMax}
    \mathcal{L}_{\rm MM}(F_{\m\n};\g)
      = S \cosh\g  + \sqrt{S^{2}+P^{2}}  \sinh\g~,
\end{align}
where 
\begin{align}\label{SP}
    S := -\tfrac{1}{4}F_{\mu\nu}F^{\mu\nu}= \tfrac{1}{2}(\vec{E}^2 - \vec{B}^2), \qquad
    P := -\tfrac{1}{4}F_{\mu\nu}\tilde{F}^{\mu\nu}= \vec{E}\cdot \vec{B},
\end{align}
are the invariants of the electromagnetic field strength $F_{\m\n}$ \cite{Minkowski}, and $\g\geq 0$ is a coupling constant.
Then, they replaced the coupling constant by a scalar field $\f$ and introduced the new model 
\begin{align}\label{RT-Lagrangian}
    \mathcal{L}(F_{\m\n};\f)
      = S \cosh\phi  + \sqrt{S^{2}+P^{2}} \sinh\phi - W(\phi),
\end{align}
where $W(\phi)$ is an arbitrary potential. This is the Russo-Townsend formulation for 
self-dual nonlinear electrodynamics. The scalar field $\f$ is auxiliary since it enters the Lagrangian without derivatives. 
Under mild conditions on the potential $W(\f)$, its equation of motion 
\bea
\frac{\pa}{\pa \f}    \mathcal{L}(F_{\m\n};\phi) =0
\eea
allows one to express $\f$ in terms of the field strength $F_{\m\n}$. Plugging the solution $\f = \f(F_{\m\n})$ back into 
\eqref{RT-Lagrangian} leads to a self-dual theory $\cL(F_{\m\n})$. 

The Russo-Townsend construction has a simple generalisation. Consider a model for self-dual nonlinear electrodynamics with its Lagrangian $L(F_{\mu\nu};g)$ depending on a duality-invariant parameter $g$, 
\begin{align}\label{selfdualequation}
    G^ {\mu \nu}\td G_{\mu \nu}  +F^{\mu \nu}\td F_{\mu \nu} =0~, \qquad 
        \tilde{G}_{\mu\nu} :=\frac{1}{2}\varepsilon_{\mu\nu\sigma\rho}G^{\sigma\rho} = 2\,\frac{\partial {L}}{\partial F^{\mu\nu}}~.
\end{align}
This self-duality equation is required for the theory to possess  invariance under $\sU(1)$ duality rotations.
Replacing the parameter $g$ in $L(F_{\mu\nu};g)$  by a duality-invariant scalar field $\j$ results in a self-dual theory $L(F_{\mu\nu};\j)$. This point was discussed long ago in the context of ${\cal N} = 1$ supersymmetric nonlinear electrodynamics \cite{KT1,KT2}.
Adding an arbitrary function, $W(\j)$, to the Lagrangian $L(F_{\mu\nu};\j)$ results in a self-dual theory,
\bea
\mathfrak{L}(F_{\mu\nu};\j) = L(F_{\mu\nu};\j) + W(\j)~.
\label{potential}
\eea
 Under reasonable conditions, the scalar field $\j$ may be integrated out, 
 using its algebraic equation of motion 
 \bea
\frac{\pa}{\pa \j}    \mathfrak{L}(F_{\m\n};\j) =0~,
\label{EoM1.7}
\eea
 and then one arrives at a new self-dual theory theory $L(F_{\mu\nu}) =  \mathfrak{L}\big(F;\j (F)\big) $. 
 Making different choices of $W(\j)$ allows one to generate different models for self-dual
nonlinear electrodynamics. Choosing a different seed Lagrangian $L(F_{\mu\nu};g)$ leads to another auxiliary-field formulation for self-dual nonlinear electrodynamics. 

In this note we will specify $L(F_{\mu\nu};g)$  to be the Born-Infeld Lagrangian \cite{Born:1934gh}
\bea
L_{\rm BI} (F_{\m\n}; g) = g-\sqrt{g^{2} -2g S -P^2 }~
\eea
and work out an example corresponding to a particular potential, eq. \eqref{WMMB}.
We will also discuss $\cN=1$ supersymmetric and other generalisations. 

%%%%%%%%%%%%%%%%%%%%%%%%%%%%%%%%%%%%%%%%%%%%%%%%%
%%%%%%%%%%%%%%%%%%%%%%%%%%%%%%%%%%%%%%%%%%%%%%%%%

\section{The model} 

Our auxiliary-field formulation for self-dual nonlinear electrodynamics is described by the following model 
\begin{equation}
{\mathfrak L}(F_{\m\n} ;\psi)= L_{\rm BI} (F_{\m\n}; \psi) +W(\psi)  
=\psi-\sqrt{\psi^{2} -2\psi S -P^{2} }+W(\psi )~.
   \label{Laux}
\end{equation}
By construction, ${\mathfrak L}(F_{\m\n} ;\psi)$ is a solution of the self-duality equation \eqref{selfdualequation}.
The equation of motion for $\j$ is 
\begin{equation}\label{eom}
1-\frac{\psi-S}{\sqrt{\psi^{2} -2\psi S -P^{2} }}+W'(\psi)=0~.
\end{equation}
The necessary condition for this equation to allows one to express $\j$ in terms of $S$ and $P$ is
\begin{equation}
\frac{S^{2}+ P^{2} }{(\psi^{2} -2\psi S - P^{2} )^{3/2} }+W''(\psi)\neq0 \label{ift}~.
\end{equation}
If $ W''(\psi)\geq0 $ then \er{ift} is the sum of two non-negative terms, so we
can say with certainty that a solution $ \psi(S,P) $ exists. As such, we
will explore potentials that satisfy this condition.

%%%%%%%%%%%%%%%%%%%%%%%%%%%%%%%%%%%

\subsection{ModMaxBorn from Born-Infeld}

Here we consider the following choice of $W(\j)$
\begin{equation}
 W_{\text{MMB}}(\psi)=T-\psi-\sqrt{\psi^{2}-2 T \psi  \cosh{\gamma}+T^{2}  } ~, 
 \label{WMMB}
\end{equation} 
where $ T $ defines the `Born-scale' and $ \gamma $ is a dimensionless
parameter. 
In this case the equation of motion \eqref{eom} is 
\begin{align}
    \frac{T\cosh{\gamma}-\psi}{\sqrt{\wmmbd}}-\frac{\psi-S}{\sqrt{\psi^{2} -2\psi S -P^{2} }}=0~. \label{mmb-eom}
\end{align}
Since
\begin{align}
    W''_{\text{MMB}}(\psi)=\frac{T^{2}\sinh^{2}\gamma }{(\wmmbd)^{3/2}  }>0 ~,
\end{align}
the condition \eqref{ift} is satisfied, and there exists a solution $\psi= \psi(S,P) $ to \er{mmb-eom}.
Solving for $ \psi(S,P) $ analytically and then substituting this solution into
\er{Laux} allows one to eliminate $\psi$. This calculation is given in Appendix \ref{WMMBProof}. 
As a result, making use of \er{WMMB} produces a new self-dual theory
\begin{align}
L^{(\s)}
     &=T-\sigma\sqrt{T^{2} -2T \big[S \cosh\gamma
     +\sigma \sqrt{S^{2}+P^{2} } \sinh \gamma \big]-P^{2}}~,
     \label{LMMBh1}
\end{align}
where $ \sigma=\pm1 $. It is an instructive exercise to check that, for each choice of $\s$, 
$L^{(\s)}$
is a solution of the self-duality equation \eqref{selfdualequation}
which is known to be equivalent to 
\begin{equation}
    P(L^2_S-L^2_P-1)=2SL_SL_P~, \label{selfdual2}
\end{equation}
with $L_S = \pa L/\pa S$ and  $L_P = \pa L/\pa P$.

The choice $ \sigma=1 $ in \eqref{LMMBh1} yields the ModMaxBorn theory \cite{Bandos:2020hgy}
\begin{equation}
    L_{\text{MMB}}\label{LMMB}(S,P)= T-\sqrt{T^{2}-2T \cL_{\text{MM}}-P^{2}}~,
\end{equation}
where
$    \cL_{\text{MM}}(S,P) $ denotes \eqref{ModMax}.
Subsequently setting $ \gamma = 0 $ recovers the Born-Infeld model.   

The choice $\s=-1$ should be discarded since $ L^{(-1)}_S$ is negative.  

%%%%%%%%%%%%%%%%%%%%%%%%%%%%%%%%%%%%

\subsection{ModMax from Born-Infeld}

As our second example we consider a linear potential 
\bea
W_{\rm MM} (\psi) = \lambda \psi~,
\eea
with $\lambda$  a dimensionless coupling constant, $\lambda >0$. The resulting model 
\bea
{\mathfrak L}_{\rm MM}(F_{\m\n} ;\psi)
=\psi-\sqrt{\psi^{2} -2\psi S -P^{2} }+\lambda \psi
\label{ModMax-new}
\eea
is conformal if $\psi$ is chosen to be a conformally primary scalar field of dimension $+4$.
Integrating out $\psi $, with the aid of its equation of motion, leads to the model 
\bea
L_{\rm MM} (S,P) = \omega S + \sqrt{w^2-1} \sqrt{S^2+P^2} ~, \qquad \o = \lambda +1~.
\label{BItoMM}
\eea
Representing 
\bea
\o=\cosh \gamma ~, \qquad \sqrt{\o^2-1} = \sinh \gamma~,
\eea
we observe that \eqref{BItoMM} coincides with the ModMax theory \eqref{ModMax}.

%%%%%%%%%%%%%%%%%%%%%%%%%%%%%%%%%%%%%%%%%%%%%
%%%%%%%%%%%%%%%%%%%%%%%%%%%%%%%%%%%%%%%%%%%%%

\section{Supersymmetric generalisations}

Now we briefly discuss supersymmetric generalisations of the  formulation proposed in the previous section.
General $\sU(1)$ duality-invariant models for supersymmetric nonlinear electrodynamics were constructed in \cite{KT1,KT2} in the rigid supersymmetric case and extended to supergravity in \cite{KMcC,KMcC2}.
They belong to the family of  nonlinear vector multiplet theories of the general 
form\footnote{We make use of the Grimm-Wess-Zumino superspace geometry \cite{GWZ} for the old minimal formulation for $\cN=1$ supergravity
\cite{WZ,old1,old2}, see \cite{BK,WB} for a review. Our superspace conventions follows \cite{KRT-M_N=1}.
In particular, $E$ is the full superspace measure, while $\mathcal{E}$ denotes the chiral density.} 
\bea
S[W,{\bar W};\U] &=&
\frac{1}{4} \int  \rd^4 x \rd^2 \q  \,\cE \, W^2 +{\rm c.c.}
\non \\
&& + \frac14  \int \rd^4 x \rd^2 \q \rd^2\bar \q \,E \,
\frac{W^2\,{\bar W}^2}{\U^2}\,
\L\left(\frac{u}{\U^2},
\frac{\bar u}{\U^2}\right)~,
\label{SUSYaction}
\eea
where $W^2 =W^\a W_\a$ and $\bar W^2 = \bar W_\ad \bar W^\ad$,  the complex variable $u$ is defined by
\bea
u  := \frac{1}{8} (\cD^2 - 4  \bar R)  W^2~,
\eea
and $\U$ is a nowhere vanishing real  scalar superfield.
This theory proves to possess $\sU(1)$ duality invariance provided the action obeys 
the $\cN=1$ self-duality equation 
\bea
{\rm Im} \int \rd^4 x \rd^2 \q  \,\cE \Big\{ W^\a W_\a  +M^\a M_\a \Big\} =0~, \qquad
{\rm i}\,M_\a := 2\, \frac{\d }{\d W^\a}\,S[W , {\bar W};\U]~,
\label{SDE1}
\eea
in which $W_\a$ is taken to be a general chiral spinor. If this equation is satisfied for $\U = {\rm const}$, it is also satisfied for an arbitrary  nowhere vanishing real  scalar superfield $\U$.

The self-duality equation \eqref{SDE1} implies an equation on the self-interaction 
$\L(\o, \bar \o)$ appearing in \eqref{SUSYaction}.  It is 
\bea
{\rm Im}\, \bigg\{ \frac{\pa (\o \, \L) }{\pa \o}
- \bar{\o}\,
\left( \frac{\pa (\o \, \L )  }{\pa \o} \right)^2 \bigg\} = 0~,
\label{GZ4}
\eea
see \cite{KT2} for the technical details.

It is worth pointing out that setting  $\U={\mathfrak g}^{-1} ={\rm const}$ in \eqref{SUSYaction} and choosing
\bea
\L_{\rm SBI}(u, \bar u) &=& \frac{{\mathfrak g}^2 }
{ 1 + \hf\, A \, +
\sqrt{1 + A +\frac{1}{4} \,B^2} }~,
\qquad  A =   {\mathfrak g}^2(u+\bar u)~, \quad 
B = {\mathfrak g}^2(u-\bar u)
\eea
defines the $\cN=1$ supersymmetric Born-Infeld action \cite{CF}.
This $\sU(1)$ duality-invariant theory is a Goldstone multiplet 
action for partial   $\cN=2 \to \cN=1$ supersymmetry 
breaking in Minkowski space \cite{BG,RT},  as well as in the following maximally supersymmetric backgrounds \cite{KT-M16}: 
(i)  ${\mathbb R} \times S^3$; (ii) ${\rm AdS}_3 \times {\mathbb R}$;
and (iii) a supersymmetric plane wave.

Given a model for nonlinear electrodynamics, 
its Lagrangian $L(F_{\m\n})$ can be expressed in terms of the 
 two independent invariants of the electromagnetic field \eqref{SP}, $L(S,P)$, 
 or equivalently $L(\o, \bar \o)$, with $\o = -S - {\rm i} P$. 
Representing $L(\o, \bar \o)$ in the form \cite{KT1,KT2}
\bea
L(\o, \bar \o)  = -\hf \, \Big( \o + \bar{\o} \Big) +
\o \, \bar{\o} \; \L (\o, \bar{\o} )~,
\label{self-interaction}
\eea
the self-duality equation \eqref{selfdualequation} turns into \eqref{GZ4}.
 The function $\L(\o, \bar \o)$ in \eqref{self-interaction} is real analytic for those self-dual theories which possess a weak-field limit.  
 Thus every self-dual nonlinear electrodynamics \eqref{self-interaction}
 has the $\cN=1$ supersymmetric extension given by \eqref{SUSYaction}, as established in  \cite{KT1}.
 For the ModMax theory \eqref{ModMax} it is \cite{K21}
\bea
\L_{\rm MM}(\o, \bar \o) = \frac{\sinh \g }{\sqrt{\o\bar \o} } 
- \hf (\cosh \g -1) \Big( \frac{1}{\o} + \frac{1}{\bar \o}\Big)~.
\eea

In earlier publications \cite{KMcC,KMcC2, K21} $\U$ was chosen to be a composite primary superfield of dimension $+2$ constructed in terms of the compensating multiplet and supersymmetric matter, for example 
\bea
\U    &=& S_0 \bar S_0 \,
{\rm exp} \Big(-\frac{1}{3} K (\vf^i,{\bar \vf}^{\bar j})\Big)~, \qquad \bar \cD_\ad S_0=0~,
\quad \bar \cD_\ad \vf^i =0~,
\eea
where $S_0$ is the chiral compensator of old minimal supergravity, $\vf^i$ matter chiral superfields, and 
$K(\vf,{\bar \vf})$ the K\"ahler potential of a K\"ahler manifold.

In this paper, we consider $\U$ to be a dynamical superfield. 
Then we can introduce a new generating formulation for $\sU(1)$ duality-invariant supersymmetric theories. 
Specifically, we fix a seed $\sU(1)$ duality-invariant model $S[W,{\bar W};\U]$, say the super Born-Infeld action, and introduce a model of the form  
\bea
{\mathfrak S}[W,{\bar W};\U]  =
S[W,{\bar W};\U]  + S[\U] ~,
\label{SUSYmodel}
\eea
for some functional $S[\U]$. This model is a solution of the $\cN=1$ self-duality equation
\bea
{\rm Im} \int \rd^4 x \rd^2 \q  \,\cE \Big\{ W^\a W_\a  +\mathfrak{M}^\a \mathfrak{M}_\a \Big\} =0~, \qquad
{\rm i}\,\mathfrak{M}_\a := 2\, \frac{\d }{\d W^\a}\,\mathfrak{S}[W , {\bar W};\U]~.
\eea
We assume that the equation of motion 
\bea
\frac{\d }{\d \U}\,\mathfrak{S}[W , {\bar W};\U] =0
\label{SUSYEoM}
\eea
allows one to express $\U$ in terms of the chiral field strength $W_\a$ and its conjugate. 
Then the action \eqref{SUSYmodel}
turns into that describing a model for self-dual supersymmetric nonlinear electrodynamics.

As a simple application of the generating formulation \eqref{SUSYmodel}, we choose
\bea
S[\U]=   \int \rd^4 x \rd^2 \q \rd^2\bar \q \,E \, f (\U)~,
\label{TrivialModel}
\eea
where $f(\U)$ is characterised by the conditions 
\bea
f'(\U_0) =0~, \qquad f''(\U_0) \neq 0~,
\eea
with $\U_0 \neq 0$ a unique solution of the equation $f'(\U)=0$. Solving the equation of motion \eqref{SUSYEoM} and plugging the solution back in \eqref{SUSYmodel} leads to the following self-dual theory: 
\bea
S &=&
\frac{1}{4} \int  \rd^4 x \rd^2 \q  \,\cE \, W^2 +{\rm c.c.}
\non \\
&& + \frac{1}{4 \U_0^2}  \int \rd^4 x \rd^2 \q \rd^2\bar \q \,E \,
{W^2\,{\bar W}^2}\,
\L\left(\frac{u}{\U_0^2},
\frac{\bar u}{\U_0^2}\right) + f(\U_0)  \int \rd^4 x \rd^2 \q \rd^2\bar \q \,E ~.
\eea
Here the last term is proportional to the supergravity action \cite{WZ}.
In order to be able to generate more interesting self-dual models,  one has to replace the function $f(\U)$ in 
\eqref{TrivialModel} with that involving spinor covariant derivatives of $\U$.

The above formulation does not work if the super ModMax theory
\cite{K21,BLST2}
\bea
S[W,{\bar W};\g ] &=&
\frac{1}{4} \cosh \g \int  \rd^4 x \rd^2 \q  \,\cE \, W^2 +{\rm c.c.}
\non \\
&& + \frac{1}{4}\sinh \g   \int \rd^4 x \rd^2 \q \rd^2\bar \q \,E \,
\frac{W^2\,{\bar W}^2}{\sqrt{u\bar u} }~.
\eea
is chosen as a seed action, since $\cosh \U$ is not chiral.

Finally, we can come back to the idea of treating $\U$ as a composite superfield 
and replace \eqref{SUSYmodel} with  a chiral  formulation
\bea
{\mathfrak S}[W,{\bar W}; \vf, {\bar \vf} ]  =
S[W,{\bar W}; \bar \vf \vf]  + S[\vf, \bar \vf ] ~,
\label{SUSYmodel-chiral}
\eea
where $\vf$ is a nowhere vanishing chiral scalar superfield, $\bar \cD_\ad \vf =0$.
We assume that the equation of motion 
\bea
\frac{\d }{\d \vf}\,\mathfrak{S}[W , {\bar W};\vf, \bar \vf ] =0
\label{SUSYEoM-chiral}
\eea
allows one to express $\vf$ and its conjugate in terms of the chiral field strengths $W_\a$ and $\bar W_\ad$. Then the action \eqref{SUSYmodel-chiral}
turns into that describing a model for self-dual supersymmetric nonlinear electrodynamics.

As an example, let us consider the following model 
\bea
{\mathfrak S}[W,{\bar W}; \vf, {\bar \vf} ]  =
S[W,{\bar W}; \bar \vf \vf]  + 
\left\{  \int  \rd^4 x \rd^2 \q  \,\cE \, f (\vf)  +{\rm c.c.}\right\}~,
\eea
where $f(\vf)$ is a holomorphic functions with the properties   
\bea
f'(\vf_0) =0~, \qquad f''(\vf_0) \neq 0~,
\eea
with $\vf_0\neq 0$ a unique solution of the equation $f'(\vf)=0$.  The equation of motion \eqref{SUSYEoM-chiral}
has a unique solution leading to the final action    
\bea
S= S[W,{\bar W}; \bar \vf_0 \vf_0 ]  + 
\left\{f (\vf_0)  \int  \rd^4 x \rd^2 \q  \,\cE    +{\rm c.c.}\right\}
\eea
which contains a supersymmetric cosmological term.  This theory is clearly self-dual.    
    
%%%%%%%%%%%%%%%%%%%%%%%%%%%%%%%%%%%%%%
%%%%%%%%%%%%%%%%%%%%%%%%%%%%%%%%%%%%%%

\section{Discussion and further generalisations}

The generating formulation for self-dual nonlinear electrodynamics proposed in this paper, which is a natural generalisation of the Russo-Townsend work \cite{Russo:2025fuc}, is more economical than the Ivanov-Zupnik approach 
 \cite{IZ_N3, IZ1, IZ2}.
 The latter makes use of an auxiliary two-form field.\footnote{To be more specific, here we refer to the so-called ``$\nu$-frame’’ version of the Ivanov-Zupnik formulation. In a recent interesting work \cite{Baglioni:2025tsc}, a relationship has been established 
between the Russo-Townsend approach and the alternative ``$\mu$-frame’’ version of the Ivanov-Zupnik formulation.} 
 However, the Ivanov-Zupnik formulation 
is truly universal in the sense that it has been extended to the followings cases: (i) ${\cal N} =1$ and ${\cal N}=2$ supersymmetric models for self-dual nonlinear electrodynamics \cite{K13,ILZ}; (ii) self-dual theories in $4n$ dimensions \cite{Kuzenko:2019nlm}; and (iii) self-dual models for $\cal N$-extended superconformal gauge multiplets \cite{KR21-2, Kuzenko:2023ebe}.

It is known that a general solution of the self-duality equation \eqref{selfdualequation} involves a real function of a real argument \cite{GR1,KT2}. Such a function naturally emerges within the Ivanov-Zupnik approach  \cite{IZ_N3, IZ1, IZ2}
as the self-interaction. A similar function originates as a scalar potential
in the Russo-Townsend formulation  \cite{Russo:2025fuc}
and its generalisation given in this paper,  eqs.  \eqref{RT-Lagrangian} and \eqref{potential}.

The equation of motion \eqref{EoM1.7} corresponding to our model \eqref{potential} has an interesting interpretation. 
Since the parameter $g$ in $L(F_{\mu\nu};g)$ is duality invariant , it is well known that $\pa L(F_{\mu\nu};g)/\pa g$ is a duality-invariant observable \cite{GZ2,GZ3}. It is also known that this observable may be expressed in terms of the energy-momentum tensor\footnote{This theorem extends  several explicit examples considered earlier in the literature in the context of $T\bar T$ deformations \cite{Conti:2018jho, Babaei-Aghbolagh:2022uij, Ferko:2022iru}.} 
\cite{Ferko:2023wyi},
\bea
\frac{\pa}{\pa g}  L(F_{\mu\nu};g) = {\mathfrak F}(T_{\m\n}; g)~.
\eea
Thus the equation of motion \eqref{EoM1.7}  can be recast in the form 
\bea
{\mathfrak F}(T_{\m\n}; \j) + W'(\j) =0~.
\eea
This equation means that the dynamics of $\j$ is determined by the energy-momentum tensor. 
A similar conclusion is expected in the case of supersymmetric self-dual systems \eqref{SUSYmodel}  
where the dynamics of $\U$ should be determined by the supercurrent computed in \cite{KMcC}. This conjecture is supported by several examples of consistent $T\bar T$ flows in $\sU(1)$ duality-invariant models for supersymmetric nonlinear electrodynamics \cite{Ferko:2019oyv, Ferko:2023ruw}.

Our construction admits a simple extension to  $\sU(1)$ 
duality-invariant nonlinear models for a gauge $(2p-1)$-form in $d=4p$ dimensions \cite{Tanii, AT, ABMZ}
(see also \cite{KT2,AFZ,Tanii2} for a review). It can also be generalised to the case of 
self-dual supersymmetric nonlinear sigma models in four dimensions \cite{Kuzenko:2013pha, Kuzenko:2023ysh}.

In conclusion, we point out that it would be interesting to study quantum aspects of the ModMax theory using its novel Born-Infeld-like reformulation \eqref{ModMax-new}. 
\\
%%%%%%%%%%%%%%%%%%%%%%%%%%%%

\noindent
{\bf Acknowledgements:}\\
We are grateful to Ian McArthur for useful comments and suggestions. 
The work of SMK is supported in part by the Australian Research Council, project DP230101629.

\appendix 

\section{Solving  the equation of motion for the auxiliary field }
\label{WMMBProof}

If we define 
\begin{align}
    A : = \axd , \quad B : = \wmmbd, \quad \kappa :  =S^{2}+P^{2}~,
\end{align} where it is understood that $ A,B,\kappa>0 $, then with \er{WMMB} the Lagrangian density in \er{Laux} may be written simply as 
\begin{equation}
    \cL=T-\sqrt{A}-\sqrt{B}~. \label{reduced-LMMB}
\end{equation}
From \er{mmb-eom}, we can also write the reduced equations of motion
\begin{align}
    \frac{T\cosh{\gamma}-\psi}{\sqrt{B}}=\frac{\psi-S}{\sqrt{A}}  \label{reduced-eom}
\end{align} 
which we can use to eliminate $ B $ from  \er{reduced-LMMB}:
\begin{align}
   \cL=T- \sqrt{A}\l(\frac{T\cosh\gamma-\psi}{\psi-S} +1 \r)\label{reduced-LMMB-2}
\end{align}
It will also be useful to note that 
\begin{align}
    A=\axd=(\psi-S)^{2}-\kappa, \label{aid}
\end{align}
and 
\begin{align}
  B=\wmmbd = (T\cosh\gamma-\psi)^{2}-T^{2}\sinh^{2}\gamma, \label{bid}
\end{align}
substitution of \er{bid} and \er{aid} into \er{reduced-eom} yields 
\begin{align}
    \kappa (T\cosh\gamma-\psi)^{2}=(\psi-S)^{2}T^{2}\sinh^{2}\gamma \label{reduced-eom-2}
\end{align}
To show that \er{reduced-LMMB-2} is indeed \er{LMMB} it suffices to show that 
\begin{align}
    \sqrt{A}\l(\frac{T\cosh\gamma-\psi}{\psi-S}+1\r)=\sqrt{T^{2}-2T \cL_{\text{MM}}-P^{2}  }.\label{LMMB-condition}
\end{align}
Squaring the LHS of \er{LMMB-condition} we get
\begin{align} 
     A\l(\frac{T\cosh\gamma-\psi}{\psi-S}+1\r)^{2}=  A\frac{(T\cosh\gamma-\psi)^{2} }{( \psi-S)^{2} }+\frac{2A(T\cosh\gamma-\psi) }{\psi-S}+A \label{A-xpr}
\end{align}
Note that the first term on the right is equivalent to $ B $ using the reduced
equations of motion in \er{reduced-eom}. 
Making use of 
\er{aid} and \er{reduced-eom-2}, for  the second term on the right of \er{A-xpr} we get
\begin{align}
    \frac{2A(T\cosh\gamma-\psi) }{\psi-S}&=2(\psi-S)(T\cosh\gamma-\psi)-\frac{2\kappa(T\cosh\gamma-\psi)}{\psi-S}\non \\
    &=2(T\psi \cosh \gamma - \psi^{2}- S T \cosh\gamma +S \psi )\mp2T (\sinh\gamma)\sqrt{\kappa}\non \\
    &=2 T\psi\cosh\gamma-2\psi^{2}- 2T[(\cosh\gamma)S\pm (\sinh \gamma)\sqrt{\kappa}]+2 S\psi~.
\end{align}
As a result \er{A-xpr} becomes
\begin{align}
        A\l(\frac{T\cosh\gamma-\psi}{\psi-S}+1\r)^{2}= T^{2} -2T[(\cosh\gamma)S\pm (\sinh \gamma)\sqrt{S^{2}+P^{2} }]-P^{2}.
\end{align}
Substituting the square root of this into the reduced Lagrangian in
\er{reduced-LMMB-2}, with $ \sigma=\pm 1 $,  one arrives at \eqref{LMMBh1}.

\begin{footnotesize}

\end{footnotesize}


\begin{thebibliography}{66}

%\cite{Russo:2025fuc}
\bibitem{Russo:2025fuc}
J.~G.~Russo and P.~K.~Townsend,
``Simplified self-dual electrodynamics,''
JHEP \textbf{10}, 120 (2025)
%doi:10.1007/JHEP10(2025)120
[arXiv:2505.08869 [hep-th]].
%4 citations counted in INSPIRE as of 13 Nov 2025



\bibitem{GZ1}
M. K.~Gaillard and B.~Zumino,
``Duality rotations for interacting fields,''
Nucl.\ Phys.\  {\bf B193},  221 (1981). 

\bibitem{BB} I. Bialynicki-Birula, ``Nonlinear electrodynamics: Variations on a theme by Born and Infeld,'' in {\it Quantum Theory of Particles and Fields}, 
B. Jancewicz and J. Lukierski (Eds.), 
World Scientific, 1983, pp. 31--48. 


\bibitem{GR1}
G.~W.~Gibbons and D.~A.~Rasheed,
``Electric-magnetic duality rotations in nonlinear electrodynamics,''
Nucl.\ Phys.\  {\bf B454}, 185 (1995) 
[arXiv:hep-th/9506035].

\bibitem{GR2}
G. W.~Gibbons and D. A.~Rasheed,
``SL(2,R) invariance of non-linear electrodynamics
coupled to an axion and a dilaton,''
Phys.\ Lett.\  {\bf B365}, 46 (1996) 
[hep-th/9509141].

\bibitem{GZ2}
M.~K.~Gaillard and B.~Zumino,
``Self-duality in nonlinear electromagnetism,''
in {\it Supersymmetry and Quantum Field Theory},
J.~Wess and V.~P.~Akulov (Eds.), Springer Verlag, 1998, pp. 121--129 [arXiv:hep-th/9705226].

\bibitem{GZ3}
M.~K.~Gaillard and B.~Zumino,
``Nonlinear electromagnetic self-duality
and Legendre transformations,'' in {\it Duality and
	Supersymmetric Theories}, D.~I.~Olive and
P.~C.~West (Eds.), Cambridge University Press,
1999, pp. 33--48 [hep-th/9712103].


%\cite{Babaei-Aghbolagh:2025uoz}
\bibitem{Babaei-Aghbolagh:2025uoz}
H.~Babaei-Aghbolagh, B.~Chen and S.~He,
``Root-$T\bar T$ flows unify 4D duality-invariant electrodynamics and 2D integrable sigma models,''
Phys. Rev. D \textbf{112}, no.10, L101702 (2025)
%doi:10.1103/1r4p-3r5q
[arXiv:2507.22808 [hep-th]].
%4 citations counted in INSPIRE as of 08 Dec 2025

\bibitem{BLST}
I.~Bandos, K.~Lechner, D.~Sorokin and P.~K.~Townsend,
``A non-linear duality-invariant conformal extension of Maxwell's equations,''
Phys. Rev. D \textbf{102}, 121703 (2020)
%doi:10.1103/PhysRevD.102.121703
[arXiv:2007.09092 [hep-th]].

\bibitem{Kosyakov}
B.~P.~Kosyakov,
``Nonlinear electrodynamics with the maximum allowable symmetries,''
Phys. Lett. B \textbf{810}, 135840 (2020)
%doi:10.1016/j.physletb.2020.135840
[arXiv:2007.13878 [hep-th]].



\bibitem{K21}
S.~M.~Kuzenko,
``Superconformal duality-invariant models and $\mathcal{N} = 4$ SYM effective action,''
JHEP \textbf{09}, 180 (2021)
[arXiv:2106.07173 [hep-th]].


\bibitem{IZ_N3} 
  E.~A.~Ivanov and B.~M.~Zupnik,
  ``N=3 supersymmetric Born-Infeld theory,''
  Nucl.\ Phys.\ B {\bf 618}, 3 (2001)
  %doi:10.1016/S0550-3213(01)00540-5
  [hep-th/0110074].

%\cite{Ivanov:2002ab}
\bibitem{IZ1} 
  E.~A.~Ivanov and B.~M.~Zupnik,
  ``New representation for Lagrangians of self-dual nonlinear electrodynamics,''
 in {\it Supersymmetries and Quantum Symmetries. Proceedings of the 16th Max Born Symposium, SQS'01: Karpacz, Poland, September 21--25, 2001}, E. Ivanov (Ed.), Dubna, 2002, pp. 235--250 
 [hep-th/0202203].
  %%CITATION = HEP-TH/0202203;%%

%\cite{Ivanov:2003uj}
\bibitem{IZ2} 
  E.~A.~Ivanov and B.~M.~Zupnik,
  ``New approach to nonlinear electrodynamics: Dualities as symmetries of interaction,''
  Phys.\ Atom.\ Nucl.\  {\bf 67}, 2188 (2004)
  [Yad.\ Fiz.\  {\bf 67}, 2212 (2004)]
  [hep-th/0303192].
  %%CITATION = HEP-TH/0303192;%%

\bibitem{Minkowski} H. Minkowski, ``Die Grundgleichungen f\"ur die elektromagnetischen Vorg\"ange in bewegten K\"orpern,'' Nachrichten der K. Gesellschaft der Wissenschaften zu G\"ottingen. Mathematisch-physikalische Klasse, 53–111 (1908); 
English Translation: ``The Fundamental equations for electromagnetic processes in moving bodies,'' 
in {\it Spacetime: Minkowski's Papers on Spacetime Physics},  V. Petkov (Ed.), Minkowski Institute Press, 2020, pp. 93--167.



\bibitem{KT1}
S.~M.~Kuzenko and S.~Theisen,
``Supersymmetric duality rotations,''
JHEP {\bf 0003}, 034 (2000)
[arXiv:hep-th/0001068].


\bibitem{KT2}
S.~M.~Kuzenko and S.~Theisen,
``Nonlinear self-duality and supersymmetry,''
Fortsch.\ Phys.\  {\bf 49}, 273 (2001) [arXiv:hep-th/0007231].


%\cite{Born:1934gh}
\bibitem{Born:1934gh}
M.~Born and L.~Infeld,
``Foundations of the new field theory,''
Proc. Roy. Soc. Lond. A \textbf{144}, no.852, 425-451 (1934).
%doi:10.1098/rspa.1934.0059
%2055 citations counted in INSPIRE as of 13 Nov 2025

\bibitem{Bandos:2020hgy}
I.~Bandos, K.~Lechner, D.~Sorokin and P.~K.~Townsend,
``On p-form gauge theories and their conformal limits,''
JHEP \textbf{03}, 022 (2021)
%doi:10.1007/JHEP03(2021)022
[arXiv:2012.09286 [hep-th]].


 
%\cite{Kuzenko:2005wh}
\bibitem{KMcC}
S.~M.~Kuzenko and S.~A.~McCarthy,
``Nonlinear self-duality and supergravity,''
JHEP {\bf 0302}, 038 (2003)
[hep-th/0212039].  

\bibitem{KMcC2}
S.~M.~Kuzenko and S.~A.~McCarthy,
``On the component structure of N=1 supersymmetric nonlinear electrodynamics,''
JHEP \textbf{05}, 012 (2005)
%doi:10.1088/1126-6708/2005/05/012
[arXiv:hep-th/0501172 [hep-th]].

\bibitem{GWZ} 
  R.~Grimm, J.~Wess and B.~Zumino,
  ``Consistency checks on the superspace formulation of supergravity,''
  Phys.\ Lett.\ B {\bf 73}, 415 (1978);
  ``A complete solution of the Bianchi identities in superspace,''
 Nucl.\ Phys.\ B {\bf 152}, 255 (1979).


\bibitem{WZ}
J.~Wess and B.~Zumino,
 ``Superfield Lagrangian for supergravity,''
 Phys.\ Lett.\  B {\bf 74}, 51 (1978).

\bibitem{old1}
K.~S.~Stelle and P.~C.~West,
``Minimal auxiliary fields for supergravity,''
Phys.\ Lett.\  B {\bf 74},  330 (1978).

\bibitem{old2}
S.~Ferrara and P.~van Nieuwenhuizen,
``The auxiliary fields of supergravity,''
Phys.\ Lett.\  B {\bf 74}, 333 (1978).


\bibitem{WB} J.~Wess and J.~Bagger,
{\it Supersymmetry and Supergravity},
Princeton University Press, Princeton, 1983 (Second Edition 1992).

\bibitem{BK} 
I.~L.~Buchbinder and S.~M.~Kuzenko,
{\it Ideas and Methods of Supersymmetry and
Supergravity or a Walk Through Superspace}, IOP, Bristol, 1995
(Revised Edition: 1998).

\bibitem{KRT-M_N=1}
S.~M.~Kuzenko, E.~S.~N.~Raptakis and G.~Tartaglino-Mazzucchelli,
``Superspace approaches to $\mathcal{N}=1$ supergravity,''
in: {\it Handbook of Quantum Gravity}, C. Bambi,  L. Modesto, I. Shapiro, I. (Eds.) Springer, Singapore (2023), https://doi:10.1007/978-981-19-3079-9\_40-1
[arXiv:2210.17088 [hep-th]].




\bibitem{CF}
S.~Cecotti and S.~Ferrara,
``Supersymmetric Born-Infeld Lagrangians,''
Phys.\ Lett.\ B {\bf 187}, 335 (1987).

\bibitem{BG}
J.~Bagger and A.~Galperin,
``A new Goldstone multiplet for partially 
broken supersymmetry,''
Phys.\ Rev.\ D {\bf 55}, 1091 (1997) 
[arXiv:hep-th/9608177].

\bibitem{RT}
M.~Ro\v{c}ek and A.~A.~Tseytlin,
``Partial breaking of global D = 4 supersymmetry, 
constrained  superfields, and 3-brane actions,''
Phys.\ Rev.\ D {\bf 59},  106001 (1999) 
[arXiv:hep-th/9811232].



\bibitem{KT-M16} 
  S.~M.~Kuzenko and G.~Tartaglino-Mazzucchelli,
  ``Nilpotent chiral superfield in N=2 supergravity and 
  partial rigid supersymmetry breaking,''
  JHEP {\bf 1603}, 092 (2016)
  %doi:10.1007/JHEP03(2016)092
  [arXiv:1512.01964 [hep-th]].


\bibitem{BLST2}
I.~Bandos,  K.~Lechner, D.~Sorokin and P.~K.~Townsend, 
``ModMax meets Susy,'' 
JHEP \textbf{10}, 031 (2021)
[arXiv:2106.07547 [hep-th]].

%\cite{Baglioni:2025tsc}
\bibitem{Baglioni:2025tsc}
N.~Baglioni, D.~Bielli, M.~Galli and G.~Tartaglino-Mazzucchelli,
``Relating auxiliary field formulations of $4d$ duality-invariant and $2d$ integrable field theories,''
[arXiv:2512.21982 [hep-th]].
%0 citations counted in INSPIRE as of 05 Jan 2026

\bibitem{K13} 
  S.~M.~Kuzenko,
  ``Duality rotations in supersymmetric nonlinear electrodynamics revisited,''
  JHEP {\bf 1303}, 153 (2013)
  [arXiv:1301.5194 [hep-th]].



\bibitem{ILZ}
E.~Ivanov, O.~Lechtenfeld and B.~Zupnik,
``Auxiliary superfields in N=1 supersymmetric self-dual electrodynamics,''
JHEP \textbf{05}, 133 (2013)
[arXiv:1303.5962 [hep-th]].

%\cite{Kuzenko:2019nlm}
\bibitem{Kuzenko:2019nlm}
S.~M.~Kuzenko,
``Manifestly duality-invariant interactions in diverse dimensions,''
Phys. Lett. B \textbf{798}, 134995 (2019)
%doi:10.1016/j.physletb.2019.134995
[arXiv:1908.04120 [hep-th]].
%10 citations counted in INSPIRE as of 25 Nov 2025

\bibitem{KR21-2}
S.~M.~Kuzenko and E.~S.~N.~Raptakis,
``Duality-invariant superconformal higher-spin models,''
Phys. Rev. D \textbf{104}, no.12, 125003 (2021)
[arXiv:2107.02001 [hep-th]].

%\cite{Kuzenko:2023ebe}
\bibitem{Kuzenko:2023ebe}
S.~M.~Kuzenko and E.~S.~N.~Raptakis,
``Self-duality for N-extended superconformal gauge multiplets,''
Nucl. Phys. B \textbf{997}, 116378 (2023)
%doi:10.1016/j.nuclphysb.2023.116378
[arXiv:2308.10660 [hep-th]].
%4 citations counted in INSPIRE as of 25 Nov 2025




%\cite{Ferko:2023wyi}
\bibitem{Ferko:2023wyi}
C.~Ferko, S.~M.~Kuzenko, L.~Smith and G.~Tartaglino-Mazzucchelli,
``Duality-invariant nonlinear electrodynamics and stress tensor flows,''
Phys. Rev. D \textbf{108}, no.10, 106021 (2023)
%doi:10.1103/PhysRevD.108.106021
[arXiv:2309.04253 [hep-th]].
%37 citations counted in INSPIRE as of 07 Dec 2025

\bibitem{Conti:2018jho}
R.~Conti, L.~Iannella, S.~Negro and R.~Tateo,
``Generalised Born-Infeld models, Lax operators and the $ \mathrm{T}\overline{\mathrm{T}} $ perturbation,''
JHEP \textbf{11} (2018), 007
%doi:10.1007/JHEP11(2018)007
[arXiv:1806.11515 [hep-th]].

%\cite{Babaei-Aghbolagh:2022uij}
\bibitem{Babaei-Aghbolagh:2022uij}
H.~Babaei-Aghbolagh, K.~B.~Velni, D.~M.~Yekta and H.~Mohammadzadeh,
``Emergence of non-linear electrodynamic theories from $T\bar T$-like deformations,''
Phys. Lett. B \textbf{829}, 137079 (2022)
%doi:10.1016/j.physletb.2022.137079
[arXiv:2202.11156 [hep-th]].
%94 citations counted in INSPIRE as of 07 Dec 2025

%\cite{Ferko:2022iru}
\bibitem{Ferko:2022iru}
C.~Ferko, L.~Smith and G.~Tartaglino-Mazzucchelli,
``On current-squared flows and ModMax theories,''
SciPost Phys. \textbf{13}, no.2, 012 (2022)
%doi:10.21468/SciPostPhys.13.2.012
[arXiv:2203.01085 [hep-th]].
%62 citations counted in INSPIRE as of 07 Dec 2025

\bibitem{Ferko:2019oyv}
C.~Ferko, H.~Jiang, S.~Sethi and G.~Tartaglino-Mazzucchelli,
``Non-linear supersymmetry and $ T\overline{T} $-like flows,''
JHEP \textbf{02} (2020), 016
%doi:10.1007/JHEP02(2020)016
[arXiv:1910.01599 [hep-th]].

\bibitem{Ferko:2023ruw}
C.~Ferko, L.~Smith and G.~Tartaglino-Mazzucchelli,
``Stress Tensor flows, birefringence in non-linear electrodynamics and supersymmetry,''
SciPost Phys. \textbf{15} (2023) no.5, 198
%doi:10.21468/SciPostPhys.15.5.198
[arXiv:2301.10411 [hep-th]].


\bibitem{Tanii}
Y.~Tanii,
{\it Introduction to supergravities in diverse dimensions},
hep-th/9802138.
%%CITATION = HEP-TH 9802138;%%

\bibitem{AT}
M.~Araki and Y.~Tanii,
``Duality symmetries in non-linear gauge theories,''
Int.\ J.\ Mod.\ Phys.\  {\bf A14}, 1139 (1999) 
[hep-th/9808029].
%%CITATION = HEP-TH 9808029;%%

\bibitem{ABMZ} 
 P.~Aschieri, D.~Brace, B.~Morariu and B.~Zumino,
 ``Nonlinear self-duality in even dimensions,''
 Nucl.\ Phys.\ B {\bf 574}, 551 (2000)
%  doi:10.1016/S0550-3213(00)00019-5
  [hep-th/9909021].

%\cite{Aschieri:2008ns}
\bibitem{AFZ}
P.~Aschieri, S.~Ferrara and B.~Zumino,
``Duality rotations in nonlinear electrodynamics and in extended supergravity,''
  Riv.\ Nuovo Cim.\  {\bf 31}, 625 (2008)
  [arXiv:0807.4039 [hep-th]].
  %%CITATION = RNCIB,031,625;%%

\bibitem{Tanii2} Y. Tanii, {\it Introduction to Supergravity}, Springer, 2014.

%\cite{Kuzenko:2013pha}
\bibitem{Kuzenko:2013pha}
S.~M.~Kuzenko and I.~N.~McArthur,
``Self-dual supersymmetric nonlinear sigma models,''
JHEP \textbf{09}, 042 (2013)
%doi:10.1007/JHEP09(2013)042
[arXiv:1306.3407 [hep-th]].
%4 citations counted in INSPIRE as of 24 Nov 2025

%\cite{Kuzenko:2023ysh}
\bibitem{Kuzenko:2023ysh}
S.~M.~Kuzenko and I.~N.~McArthur,
``A supersymmetric nonlinear sigma model analogue of the ModMax theory,''
JHEP \textbf{05}, 127 (2023)
%doi:10.1007/JHEP05(2023)127
[arXiv:2303.15139 [hep-th]].
%11 citations counted in INSPIRE as of 24 Nov 2025


\end{thebibliography}
\end{document}